# Thermally-activated precipitation strengthening

Guangpeng Sun, Liqiang zhang*, Bin Wen*

State Key Laboratory of Metastable Materials Science and Technology, Yanshan University, Qinhuangdao 066004, China.

**Abstract** Precipitation strengthening is a key strengthening method for metallic materials. However, the temperature effect on precipitation strengthening is still unclear to date. Based on dislocation theory, a thermally-activated precipitation strengthening model is built by considering the competition between shear and bypass mechanisms. For medium-sized precipitate particles, the thermally-activated shear mechanism dominates the precipitation strengthening, resulting in a plateau region. While, for large or very fine precipitate particles, the thermally-activated bypass mechanism dominates the precipitation strengthening, leading to the strengthening or weakening regions. Moreover, the effects of precipitate phase volume fraction, temperature, shear modulus, strain rate, and mobile dislocation density on precipitation strengthening are also investigated. This study not only provides new insights into precipitation strengthening from the perspective of thermal activation but also offers clear guidance for the design of new materials.

**Keywords:** Precipitation strengthening, thermal activation, precipitate diameter, precipitate phase volume fraction, shear modulus

* Corresponding author. E-mail address: wenbin@ysu.edu.cn
* Corresponding author. E-mail address: lqzhang@ysu.edu.cn

1. **Introduction**

   Material strengthening is an important topic in the field of structural materials research [1-3]. As a well-known material strengthening method, precipitation strengthening benefits from the dispersed precipitate phase particles blocking dislocation gliding [4, 5]. Compared to the other strengthening methods (i.e. solution strengthening, dislocation strengthening, and grain boundary strengthening [6]), precipitation strengthening can still be effective even at elevated temperatures. Therefore, precipitation strengthening, as the main method to improve the high-temperature properties of alloys, has been widely studied [7-10].

   Based on the interaction between dislocation and precipitate phase particles, the precipitation strengthening mechanism can be classified into two types: the shear mechanism and the bypass mechanism [3, 4]. To conduct a quantitative analysis of the precipitation strengthening, many models have been developed based on these two types of mechanism [4, 5, 11]. For the shear mechanism, the ordered strengthening models [12, 13] indicate that the antiphase boundary surface dominates precipitation strengthening. The modulus strengthening models [14, 15] and coherent strengthening models [16, 17] highlight that the shear modulus mismatch and the strain-field interactions also play a certain role in precipitate strengthening. For the bypass mechanism (i.e. Orowan mechanism), a dislocation overcomes the obstruction of impenetrable precipitate phase particle by leaving a dislocation loop. In this mechanism, the Orowan models emphasize that the strengthening effect is determined by the precipitate phase volume fraction, matrix shear modulus [18-20], and precipitate phase particles geometric properties (i.e. diameter, shape, and distribution) *et al.* [13, 21-23]. However, all of these models are independent of temperature, and the thermally-activated precipitation strengthening model has not been studied so far. Very recently, a thermally-activated Orowan strengthening mechanism was investigated by our group [24]. Our results indicated that under an elevated temperature condition, the thermal activation effect causes significant changes in the Orowan strengthening stress and it cannot be ignored even at room temperature. Nonetheless, it is still uncertain how the thermal activation affects the precipitation strengthening because the effect of thermal activation on shear mechanism and the effect of thermal activation on the competition between shear mechanism and bypass mechanism have not been studied.

To study the thermal activation effect on precipitation strengthening, the physical processes of dislocation shearing and bypassing precipitate phase particles are re-examined in this work. By incorporating dislocation self-interaction, competition between shear mechanism and bypass mechanism, and temperature effect, a thermally-activated precipitation strengthening model is built. Our research results indicate that the thermal activation effect has a certain impact on the shear mechanism. The precipitate phase particle diameter and thermal activation play an important role in the competition between shear mechanism and bypass mechanism. This study not only provides new insights into precipitation strengthening from the perspective of thermal activation but also offer a guidance for the design of new materials.

## 2. Mathematical modeling for thermally-activated precipitation strengthening

The interaction between the precipitate phase particles and dislocation are depicted in Fig. 1. There are primarily two mechanisms used to describe the interaction [15, 18, 25]. One is the Orowan mechanism (see Figs. 1a-1c), and the other is the shear mechanism (see Figs. 1d-1f). If the thermal activation effect is considered, the shear mechanism and bypass mechanism have a certain probability of occurrence. In this section, the thermally-activated processes for the two mechanisms are derived separately. Subsequently, the competition between the two mechanisms is explored, and a thermally-activated precipitation strengthening model is established.

## 2.1 Mathematical modeling for thermally-activated bypass mechanism

As schematically illustrated in Fig. 1a, when a moving dislocation is hindered by hard precipitate phase particles, the dislocation bows out between them. Considering the interaction between dislocation branches, the bowing dislocation is usually not in standard circular form. Under a constant shear stress ($\tau$) condition, the equilibrium configuration needs to satisfy the static stress equilibrium equation [18]:

$$\tau + \tau_{self} = 0, \tag{1}$$

where $\tau_{self}$ is the self-stress [26]. A numerical approach can be used to relax the bowing dislocation into its equilibrium position. Initially, the bowing dislocation needs to be discretized into a set of elements. Subsequently, the self-stress on each junction point can be calculated with reference to Ref. [26]. Thereafter, judge whether the net unbalanced stress, $\Delta\tau = \tau + \tau_{self}$, is less than a certain threshold (represented by '$1\% \cdot \tau$'). If this is true for all junction points, the bowing dislocation is considered to be in its equilibrium position. Otherwise, the curvature of each element needs to be relaxed through the iterative formula [26]:

$$k_{iter} = k\left[1 - \Delta\tau / \tau_{arc}\right], \tag{2}$$

where $k$ and $k_{iter}$ are the curvatures before and after the iteration, respectively. $\tau_{arc}$ is the self-stress calculated from the elements on both sides of the junction point. Repeat the process of relaxation and judging until all junction points satisfy Eq. 1; the equilibrium configuration can be obtained.

When the bowing dislocation is in its equilibrium position, the dislocation system is in a stable state. According to the dislocation theory [6], the dislocation system energy can be written as [24]:

$$\Phi_m(\tau) = E - \left(\frac{A_1 G_m b_m^2 L}{4\pi} + \frac{L+D}{LD} \cdot \frac{A_2 G_m b_m^2 y^2}{8\pi}\right)\ln\left(\frac{R_0}{r_0}\right) - A\tau b_m, \tag{3}$$

where $E = \sum \Gamma l$ is elastic energy of the equilibrium configuration, $\Gamma = \frac{G_m b_m (1-\nu \cos^2 \gamma)}{4\pi(1-\nu)} \ln\left(\frac{R_0}{r_0}\right)$ is line tension, $A_1 = \cos^2 \gamma + \frac{\sin^2 \gamma}{1-\nu}$, $A_2 = \sin^2 \gamma + \frac{\cos^2 \gamma}{1-\nu}$, $\gamma$ is angle between discrete element and Burgers vector, $\nu$ is Poisson's ratio, $l$ is arc length of the discrete element, $b_m$ and $G_m$ are magnitude of Burgers vector and shear modulus of matrix, $R_0$ is integral range of dislocation, $r_0$ is radius of dislocation core, $D$ and $L$ are precipitate diameter and interprecipitate spacing, $y$ and $A$ are maximum height and sweeping area of the equilibrium configuration.

As shear stress increases, a sequence of equilibrium configurations form a bowing path, bounded by the original straight line and critical configuration (see Fig. 1b). When the dislocation slips along this bowing path, the dislocation system energy exhibits the minimum energy pathway (MEP). Under a constant shear stress, the dislocation system energy initially decreases, followed by an increase, and ultimately reaches its maximum value at the critical configuration (see Fig. 2a). The activation energy can be expressed by the difference between the maximum and minimum values, as follows:

$$\Delta \Phi_m(\tau) = \Phi_m^{cri}(\tau) - \Phi_m^{equ}(\tau), \tag{4}$$

where $\Phi_m^{cri}(\tau)$ and $\Phi_m^{equ}(\tau)$ are the dislocation system energy when a dislocation is in its critical and equilibrium position. As the shear stress increases, the activation energy decreases (see Fig. 2c).

Temperature not only influences the mechanical activation through the shear modulus but also impacts the thermal activation effect [24]. If the thermal activation energy is sufficient to cause a dislocation to exceed the critical configuration, it triggers a thermally-activated bypass process (see Fig. 1c). The Harmonic Transition State Theory provides a methodology to predict the thermally-activated mechanism [27]. It connects the strain rate to the activation energy and temperature using the Arrhenius equation, as follows [28]:

$$\dot{\varepsilon}_m = \dot{\varepsilon}_0 \exp\left(-\Delta\Phi_m(\tau)/(k_B T)\right), \tag{5}$$

where $\dot{\varepsilon}_0 = \rho b \omega \lambda$ [29], $\rho$ is the mobile dislocations density of matrix, $\lambda$ is the mean free path of dislocations, $\omega$ and $k_B$ are Debye frequency and Boltzmann constant. The Eq. 5 indicates that when a combination of temperature and shear stress is given, there is a possibility for dislocations to bypass precipitate phase particles, resulting in a thermally-activated strain rate $\dot{\varepsilon}_m$.

**2.2 Mathematical modeling for thermally-activated shear mechanism**

When a moving dislocation is hindered by soft precipitate phase, the dislocation will overcome the particles through shear mechanism. As shown in Fig. 1d, a moving dislocation forms a kink-pair inside the precipitate phase. Due to the well-defined configuration of the kink-pair,etimes the dislocation system energy can be written as [30, 31]:

$$\Phi_p(\tau) = \frac{A_1 G_p b_p^2 h}{2\pi} \ln\left(\frac{R_0}{r_0}\right) - \frac{A_3 G_p b_p^2 h^2}{8\pi x} - hx\tau b_p, \tag{6}$$

where $x$ is kink-pair width, $h$ is kink height, $A_3 = \dfrac{(1+\nu)\cos^2\gamma + (1-2\nu)\sin^2\gamma}{1-\nu}$, $b_p$ and $G_p$ are magnitude of Burgers vector and shear modulus for precipitate phase. By inputting an increasing kink-pair widths into Eq. 6, the MEP initially increases and then decreases (see Fig. 2b).

Mathematically, taking the derivative of Eq. 6, the critical kink-pair width can be expressed as [31]:

$$x_c = \left(\frac{A_3 h b_p}{8\pi} \frac{G_p}{\tau}\right). \tag{7}$$

Therefore, the kinking path can be defined, where the kink-pair width is within a range of 0 to $x_c$ (see Fig. 1e). After substituting Eq. 7 into Eq. 6, the activation energy has an analytical formula, as follows:

$$\Delta\Phi_p(\tau) = \frac{A_1 G_p b_p^2 h}{2\pi} \ln\left(\frac{x_c}{r_0}\right) - \left(hb_p\right)^{3/2}\left(\frac{A_3 G_p \tau}{2\pi}\right)^{1/2}. \tag{8}$$

As shown in Fig. 2d, the activation energy decreases with an increasing in shear stress.

Considering temperature effect, thermal activation promotes the dislocation to shear the precipitate phase (see Fig. 1f). Therefore, the thermally-activated shearing strain rate, $\dot{\varepsilon}_p$, can be expressed as:

$$\dot{\varepsilon}_p = \dot{\varepsilon}_0 \exp\left(-\Delta\Phi_p(\tau)/(k_B T)\right). \tag{9}$$

What is noteworthy is that in Eq. 9, $\dot{\varepsilon}_0 = \rho_p b_p \omega_p \lambda_p$, where $\rho_p$, $b_p$, $\omega_p$, and $\lambda_p$ respectively represent the mobile dislocation density, Burgers vector, Debye frequency, and the mean free path for the precipitate phase materials.

## 2.3 The competition between bypass mechanism and shea mechanism

Under any combination of temperature and shear stress, there exists a probability for the dislocation bypassing or shearing the precipitate phase particles. Since both mechanisms occur concurrently and are independent, they have to compete. In addition, the combined strain rates of these two mechanisms constitute the experimentally measured strain rate. Therefore, the experimentally measured strain rate can be expressed as:

$$\dot{\varepsilon} = \dot{\varepsilon}_p + \dot{\varepsilon}_m. \tag{10}$$

The shear stress that satisfies Eq. 10 is the thermally-activated precipitate strengthening stress. In this section, a numerical method to obtain the thermally-activated precipitate strengthening stress is described, as shown in Fig. 3. Before proceeding with the solution, it is necessary to initialize the precipitate phase particle geometrical parameters ($D, L$), material properties ($G_m, b_m, G_p, b_p$), experimental deformation conditions ($\dot{\varepsilon}, T$), and a small shear stress ($\tau$). Initially, the activation energy and thermally-activated strain rate for the two mechanisms are calculated using Eq. 4, Eq. 5, and Eq. 8, Eq. 9. Subsequently, determine whether Eq. 10 is satisfied. If it is satisfied, the shear stress is the thermally-activated precipitation strengthening stress. Otherwise, increasing the shear stress, the activation energy and thermally-activated strain rate are recalculated. Repeat this operation until Eq. 10 is satisfied, the finally $\tau$ is the thermally-activated precipitation strengthening stress.

**2.4 Precipitation strengthening stress formula**

Regardless of how dislocations overcome precipitate phase particles, the softer mechanism will dominate the plastic deformation. Consequently, the precipitation strengthening stress formula can be expressed in terms of the competition between thermally-activated bypass stress and shear stress. To express the thermally-activated stress, the knowledge of activation energy for the two mechanisms is needed [32]. The phenomenological activation energy expression proposed by Kocks *et al.* [33] is used, as follows:

$$\Delta\Phi(\tau) = gGb^3 \left(1-\left(\tau/\hat{\tau}\right)^p\right)^q, \tag{11}$$

where $gGb^3$ is the shear stress-free activation energy, $\hat{\tau}$ is the critical resolved shear stress (CRSS) for bypass mechanism or shear mechanism [18, 31], $p$ and $q$ are the parameters determining the shape of the activation energy curve. After substituting Eq. 11 into Eq. 5 and Eq. 9, the thermally-activated bypass stress or shear stress can be expressed as:

$$\tau = (1-\alpha)\hat{\tau}, \tag{12}$$

where $\alpha = 1-\left(1-\left[1-k_B T \ln\left(\rho b\lambda\omega/\dot{\varepsilon}\right)/\left(gGb^3\right)^{1/q}\right]\right)^{1/p}$ is the thermal activation factor. According to our previous research [24], for the bypass mechanism, the parameters can be determined as $p=1, q=3/2$, $g = \ln D \ln L \ln(DL/(D+L))(1-\nu\cos^2\gamma)/(2\pi-2\pi\nu)$. For the shear mechanism, the parameters is $p=3/2$, $q=1$, $g=1.57$ [34]. Based on the analysis above, thermally-activated bypass stress or shear stress can be expressed through the coupling of thermal activation component and mechanical activation component, and precipitation strengthening stress is determined by the softer mechanism.

## 3. Verification of thermally-activated precipitation strengthening model

To verify the reliability of thermally-activated precipitation strengthening model, the precipitation strengthening stress predicted by this model are compared with experimental data. The calculation parameters are shown in Table 1. Fig. 4 displays the calculated precipitation strengthening stress for AlSc [35], MgAlScZr [36], FrCoNiAlV [37] and AlScZr [13] alloys as compared with the experimental data, which are highly consistent.

From a competitive standpoint, our model can also predict the changing trend of precipitation strengthening stress. Fig. 5a shows that at a constant precipitate phase volume fraction ($c$), a decrease in precipitate diameter leads to a reduction in interprecipitate spacing (refer to $L = \left(\sqrt[3]{\pi/(6c)} - 1\right)D$). For the shear mechanism, according to the previous research [31, 39], the CRSS is independent of the precipitate diameter:

$$\hat{\tau}_p = \frac{2\pi}{A_3 h^3 b^3 G} \left[\frac{A_1 G b^2 h}{2\pi} \ln\left(\frac{x_c}{r_0}\right)\right]^2 / \sqrt[3]{\pi/(6c)}. \tag{13}$$

And the $\alpha$ remains constant as the precipitate diameter changing (see Fig. 5b). Therefore, according to Eq. 12, the thermally-activated shear stress is also independent of the precipitate diameter (see Fig. 5c).

For the bypass mechanism, the CRSS increases as the precipitate diameter decreases [18]:

$$\hat{\tau}_m = \left(\frac{\ln D}{\ln L}\right)^{3/2} \frac{Gb}{\left(\sqrt[3]{\pi/(6c)} - 1\right)D} \frac{\ln L}{2\pi}. \tag{14}$$

Meanwhile, the $\alpha$ also increases with a decrease of the precipitate diameter (see Fig. 5b). According to Eq. 12, the thermally-activated bypass stress increases first and then decreases (see Fig. 5c). The reason for this phenomenon is that, with the decrease in precipitate diameter, the thermal activation contribution gradually takes a dominant role in the bypass process, leading to a decrease in the mechanical activation contribution. Therefore, the strengthening effect is weakened; for extremely fine precipitates, the stress may even decline.

Considering the competitive relationship, the precipitation strengthening stress is determined by the smaller of the two mechanisms. Consequently, the precipitation strengthening stress curve exhibits segmentation behavior. Specifically, as the precipitate diameter decreases, the curve displays distinct strengthening, plateau, and weakening regions. Because the experimental strain rate is equal to the sum of the strain rates associated with the bypass and shear mechanism (Eq. 10), according to Fig. 5d, these regions are governed by the bypass mechanism ($\geq 4$ nm), the shear mechanism (1~2 nm), and the combined bypass+shear mechanism (0~1 nm or 2~4 nm), respectively.

Based on the above analysis, it is observed that during plateau region and strengthening region, our model shares the same viewpoint with the traditional order strengthening model and Orowan strengthening model [5-7]. Namely, the plateau region is dominated by the shear mechanism, and the strengthening region is dominated by the bypass mechanism. However, in the weakening region, the traditional approach is to employ coherent+modulus strengthening models to explain this phenomenon [5-7]. Here, a new perspective is provided that the weakening region is attributed to the strong thermal activation effect, and this viewpoint is supported by existing experimental strength data [13, 35-37].

## 4. Discussion

According to thermally-activated precipitation strengthening model, the trend of precipitation strengthening stress can be readily explained. Furthermore, the influence of temperature, shear modulus, precipitate phase volume fraction, dislocation density, and strain rate on precipitation strengthening is also examined in this section.

### 4.1 Effect of the temperature and precipitate diameter on precipitation strengthening stress and thermal activation effect

In this study, the precipitation strengthening stress at different temperatures is predicted. The AlSc alloy comprises $Al_3Sc$ phases that are uniformly dispersed in the Al matrix [35]. The volume fraction of $Al_3Sc$ precipitate remains stable across a wide temperature range [40]. Concerning the temperature, the shear modulus is represented by temperature-dependent parameters $G(T)$, which can be expressed by $\partial G_{Al}/\partial T = -13.6\,\mathrm{MPa \cdot K^{-1}}$, $\partial G_{Al_3Sc}/\partial T = -26\,\mathrm{MPa \cdot K^{-1}}$. The temperature range is 50~700 K, and other parameters of AlSc alloy are shown in Table 1.

As expected, with a reduction in precipitate diameter, the precipitation strengthening stress exhibits an initial increase, followed by a decrease. Additionally, the precipitation strengthening stress decreases with rising temperature (see Fig. 6a). Based on previous analysis, the plateau region represents the shear mechanism, and noticeable temperature sensitivity is observed. Specifically, at low temperatures (50~200 K), the bypass mechanism dominates plastic deformation (see Fig. 6b-6d), thus no plateau appears on the precipitation strengthening stress curve. While at high temperatures (300~700 K), the dominant mechanism transitions from the bypass mechanism to the shear mechanism at a specific precipitate diameter (see Fig. 6e). Naturally, a plateau appears on the curve at this particular precipitate diameter. Therefore, with the increase in temperature, a plateau generally appears on the precipitation strengthening stress curve. This plateau represents the maximum strengthening effect of the material at the given temperature.

The influence of temperature and precipitate diameter on the thermal activation effect is predicted. Fig. 6f and 6g depict $\alpha$ as a function of temperature and precipitate diameter for the bypass and shear mechanisms. Regarding temperature, $\alpha$ increases with rising temperature for the two mechanisms, indicating that the thermal activation contribution rises with increasing temperature. Regarding precipitate diameter, $\alpha$ remains unchanged with varying precipitate diameter for the shear mechanism, indicating that its thermal activation contribution is independent of precipitate diameter. However, for the bypass mechanism, $\alpha$ decreases with rising precipitate diameter, indicating that the thermal activation contribution decreases with increasing the precipitate diameter.

**4.2 Effect of shear modulus on precipitation strengthening stress and thermal activation effect**

Since the precipitate diameter and the shear modulus determine how dislocations interact with precipitate particles, they are also important factors influencing precipitation strengthening [11]. To account for this influence, the precipitation strengthening stress as a function of the precipitate diameter is predicted under different shear moduli. Here, parameters $\dot{\varepsilon} = 3 \times 10^{-2} \, \text{s}^{-1}$, $b_m = 0.286$ nm, $\rho_m = 10^{13}$ m$^{-2}$, $G_m = 24 \sim 54$ GPa, $\nu_m = 0.34$, $\lambda_m = 100b$, $c = 0.75\%$, $T = 300$ K, $b_p = 0.289$ nm, $\rho_p = 10^{16}$ m$^{-2}$, $G_p = 34 \sim 64$ GPa, $\nu_p = 0.17$, $\lambda_p = 10b$ are selected.

Fig. 7a shows that precipitate phases with higher shear moduli will elevate the plateau but will not affect the Orowan strengthening stage. On the contrary, increasing the shear modulus of the matrix material will elevate the Orowan strengthening stage but will not affect the plateau, as shown in Fig. 7b. The influence of shear modulus on thermal activation effect is also predicted. As shown in Fig. 7c and 7d, $\alpha$ decreases with rising shear modulus for both mechanisms, indicating that the thermal activation contribution reduces with increasing shear modulus.

**4.3 Effect of volume fraction on precipitation strengthening stress and thermal activation effect**

The precipitate phase volume fraction also plays an important role in precipitation strengthening. To account for this influence, the precipitation strengthening stress is predicted as a function of precipitate diameter under different $c$. In this calculation, the calculation parameters of AlSc are shown in Table 1, with the precipitation phase volume fraction set within the range of 0.5%~10%.

Fig. 8a shows that the precipitation strengthening effect becomes more pronounced with the increase in precipitate phase volume fraction. This is because a higher precipitate phase volume fraction leads to a higher density of precipitate phase particles, which, in turn, results in an increased CRSS for the bypass mechanism (Eq. 14) and shear mechanism (Eq. 13). In addition, the influence of precipitate phase volume fraction on thermal activation effect is studied. As shown in Fig. 8b, $\alpha$ increases with rising precipitate phase volume fraction for bypass mechanism, indicating that the thermal activation contribution rises with increasing precipitate phase volume fraction. It's worth noting that, for the shear mechanism, the thermal activation contribution is independent of the precipitate phase volume fraction.

**4.4 Effect of matrix mobile dislocation density and strain rate on precipitation strengthening stress and thermal activation effect**

From the perspective of thermal activation analysis [34, 41], the strain rate and matrix mobile dislocation density also influence the precipitation strengthening. To account for these influence, the precipitation strengthening stress is predicted under different strain rate and mobile dislocation density. Taking AlSc as the research object, when studying the influence of matrix mobile dislocation density on precipitation strengthening, the dislocation density is set within the range of $10^8 \sim 10^{16}$ m$^{-2}$. Additionally, when studying the effect of strain rate on precipitation strengthening, the strain rate is set within the range of $0.0003 \sim 3$ s$^{-1}$.

Fig. 9a demonstrates that an increase in the mobile dislocation density does not affect the plateau, but it slightly reduces the Orowan strengthening stage. In contrast to the mobile dislocation density, elevated strain rates enhance the overall precipitation strengthening, as shown in Fig. 9b. The influence of strain rate/mobile dislocation density on thermal activation effect is predicted (see Fig. 9c and 9d). An increase in mobile dislocation density leads to enhanced $\alpha$, indicating that the thermal activation contribution rises with increasing matrix mobile dislocation density. Conversely, an increase in strain rate leads to a reduction in $\alpha$. Consequently, the thermal activation contribution decreases.

**5. The competition between the thermally-activated bypass and shear mechanism**

Based on the above analysis, numerous factors were found to affect precipitation strengthening, especially the precipitate diameter. Fig. 10 presents a schematic representation of a competition map between the thermally-activated bypass and shear mechanism. For very fine precipitate particles, due to the strong thermal activation effect, the material exhibits a weakening behavior dominated by the bypass mechanism. For medium-sized precipitate particles, as the thermally-activated shear stress is lower than the bypass stress, the material exhibits a plateau phenomenon dominated by the shear mechanism. For large precipitate particles, the material exhibits strengthening behavior dominated by the bypass mechanism, since the thermally-activated bypass stress is significantly smaller than that of the shear stress. In addition, in the weakening and strengthening regions, the thermal activation contribution significantly increases with a decrease in precipitate diameter. However, for the plateau region, the thermal activation contribution is independent of the precipitate diameter.

In addition to the precipitate diameter, the shear modulus and precipitate phase volume fraction also significantly influence the precipitation strengthening. Specifically, increasing the shear modulus not only enhances the resistance to dislocation movement but also reduces the thermal activation contribution, thereby achieving a better strengthening outcome. In addition, increasing the precipitate phase volume fraction significantly enhances the density of precipitate particles, leading to a substantial increase in the CRSS, improving the precipitate strengthening. Therefore, when designing high-temperature alloys with excellent performance, we recommend maximizing the shear modulus, precipitate phase volume fraction and setting the precipitate diameter within the moderate size range.

## 6. Conclusions

Herein, a thermally-activated precipitation strengthening model is proposed based on the competition between the bypass and shear mechanisms. The rationality and reliability of the model are verified by comparing the calculation results with experimental data for several Al alloys and steel. The results indicate that precipitate diameter considerably affects the competition between the bypass and shear mechanisms. For medium-sized precipitate particles, the thermally-activated shear mechanism dominates the precipitation strengthening, resulting in a plateau region. While for large or very fine precipitate diameter, the thermally-activated bypass mechanism dominates the precipitation strengthening, resulting in a strengthening or weakening region. This study not only provides new insights into precipitation strengthening from the perspective of thermal activation but also offers clear guidance for the design of new materials.

**Acknowledgment:** This work was supported by the National Natural Science Foundation of China (Grant Nos. 51925105, 51771165) and the National Key R&D Program of China (YS2018YFA070119).

**Author contributions**：Bin Wen conceived the project. Guangpeng Sun and Bin Wen performed the model construction and calculations. Guangpeng Sun, Liqiang zhang , Bin Wen wrote this paper.

**Conflict of interest:** The authors declare that they have no conflict of interest.

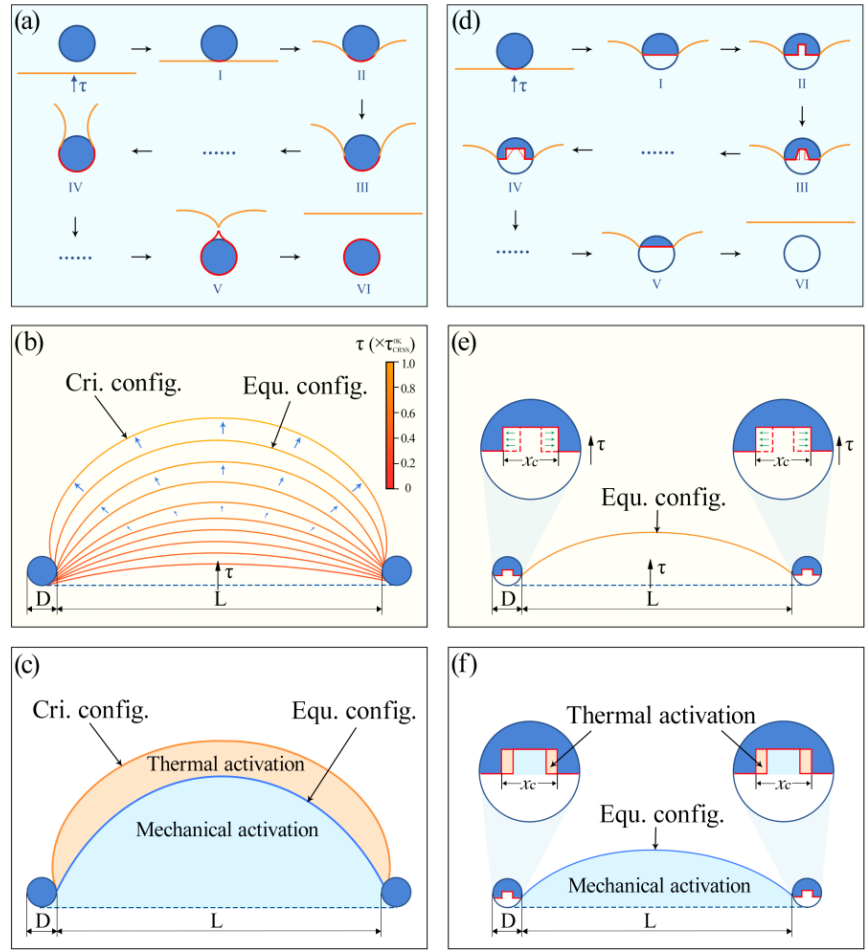

**Figure 1. Schematic diagram for the thermally-activated precipitation strengthening.** (a) The process of dislocations bypassing the precipitate phase particles. (b) The bowing path. (c) Schematic diagram of the thermally-activated bypass mechanism. (d) The process of dislocation shearing the precipitate phase particles. (e) The kinking path. (f) Schematic diagram of the thermally-activated shearing mechanism.

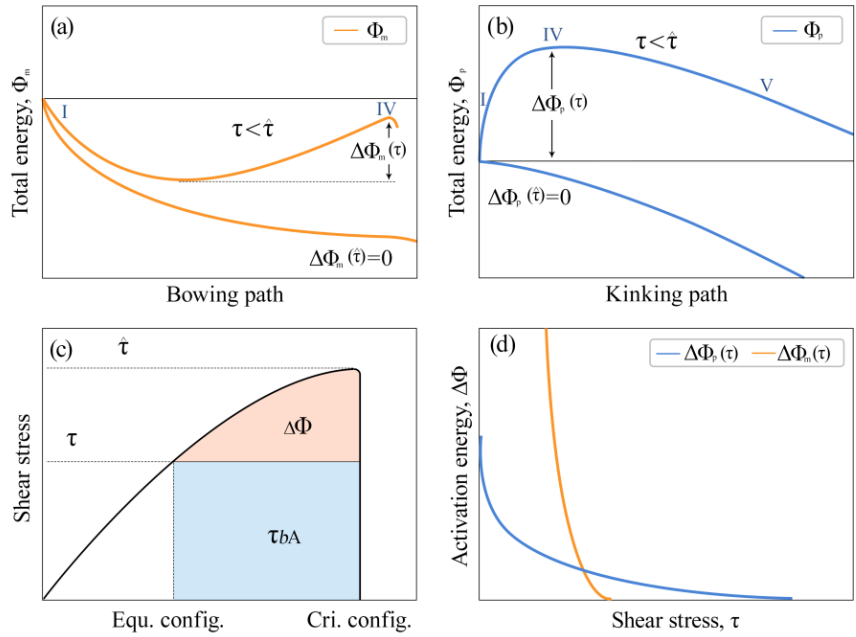

**Figure 2. Total energy and activation energy.** (a) The total energy variation with respect to the bowing path for bypass mechanism. (b) The total energy variation with respect to the kinking path for shear mechanism. (c) The contributions of thermal and mechanical activation during the process of dislocation overcoming precipitates. (d) Activation energy as a function of shear stress for bypass and shear mechanisms.

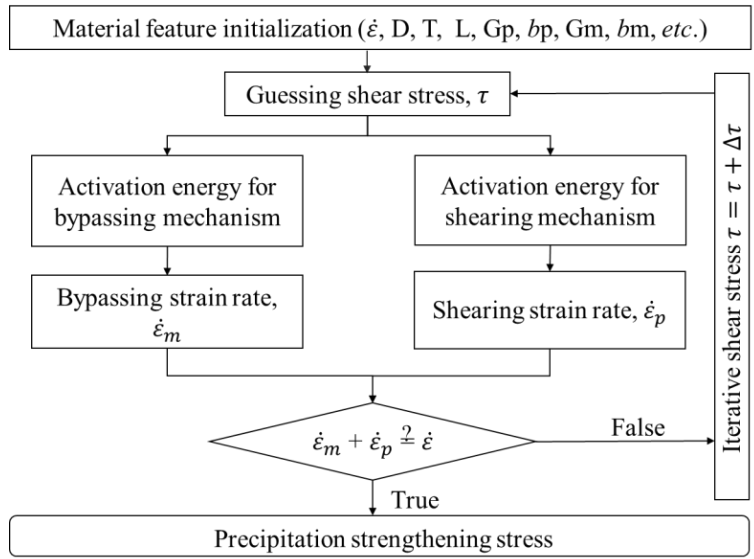

**Figure 3. The calculation method of precipitation strengthening stress.**

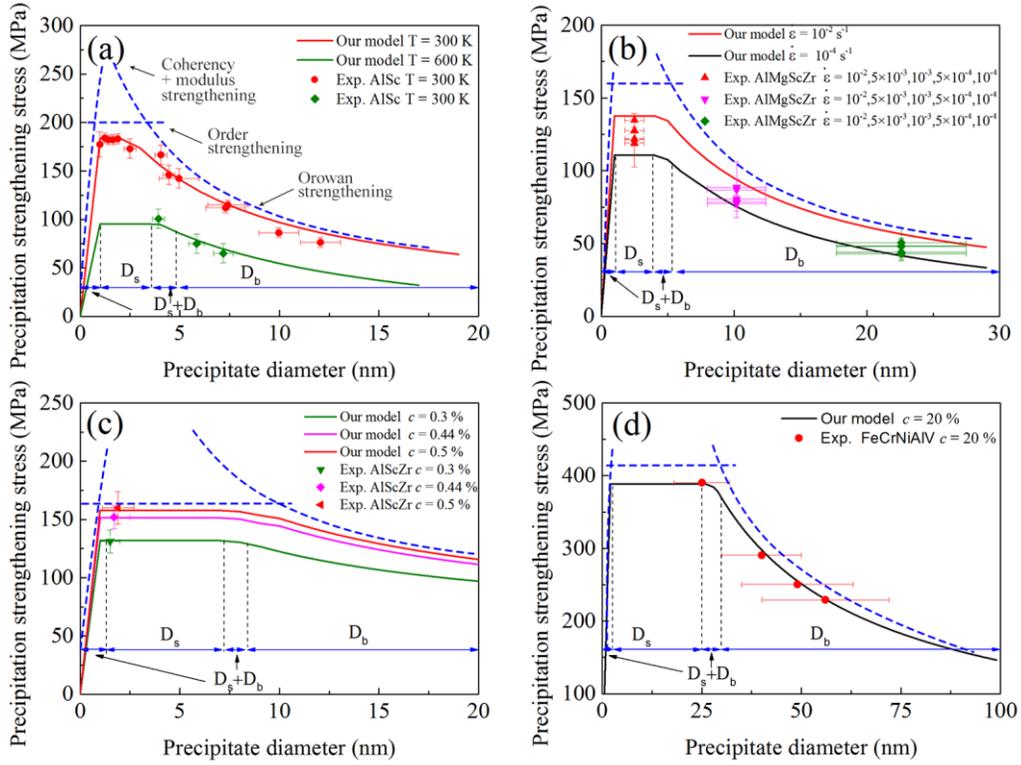

**Figure 4. Calculated precipitation strengthening stress compared with experimental data.** (a-c) Comparison between the calculated precipitation strengthening stress and experimental results for the Al alloy. The blue dashed lines are calculated from other theoretical models [5-7]. (d) Comparison of precipitation strengthening stress for steel with experimental results [37].

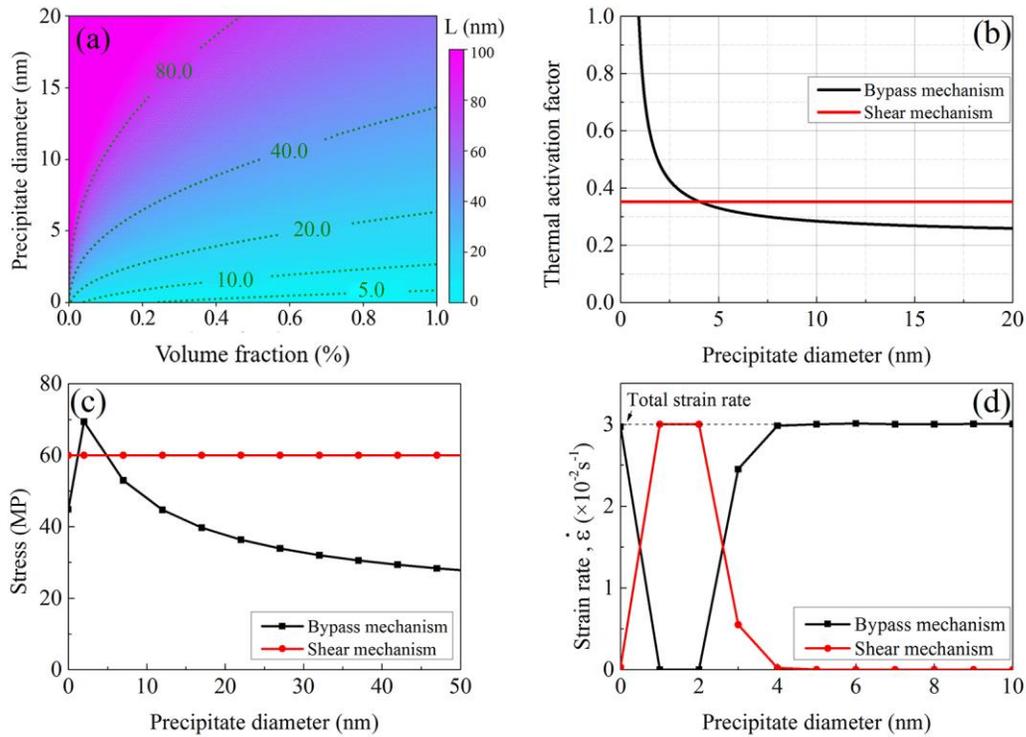

**Figure 5. Effect of precipitate diameter on deformation mechanism in AlSc alloy.** (a) The interprecipitate spacing as a function of precipitate diameter and volume fraction. (b) Effect of the precipitate diameter on thermal activation factor. (c) Thermally-activated bypass stress and shear stress as functions of precipitate diameter. (d) The strain rate varies with the precipitate diameter.

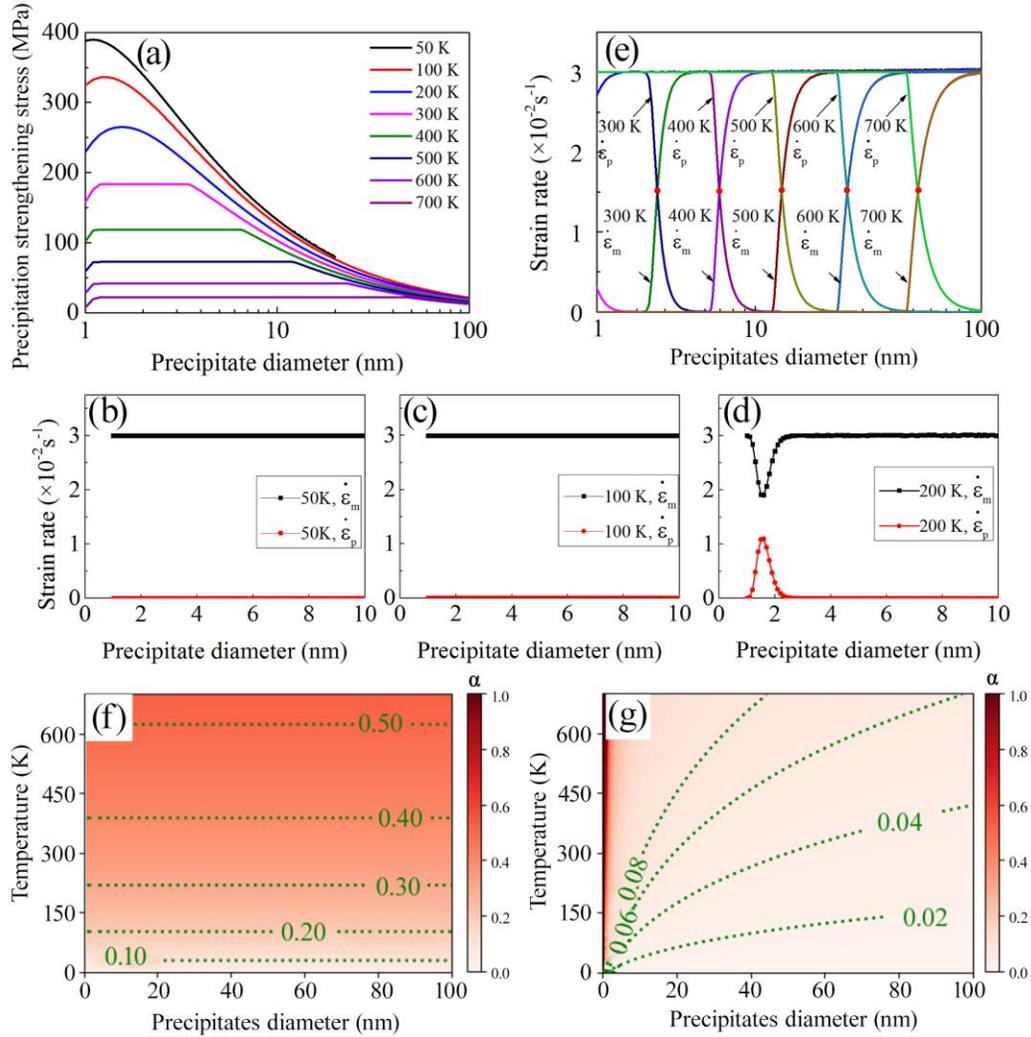

**Figure 6. Effect of temperature and precipitate diameter on precipitation strengthening.** (a) The variation of precipitation strengthening stress with precipitate diameter under different temperature. (b-e) The effects of temperature and precipitate diameter on deformation mechanism. (f-g) Effect of precipitate diameter and temperature on thermal activation factor for shear mechanism and bypass mechanism.

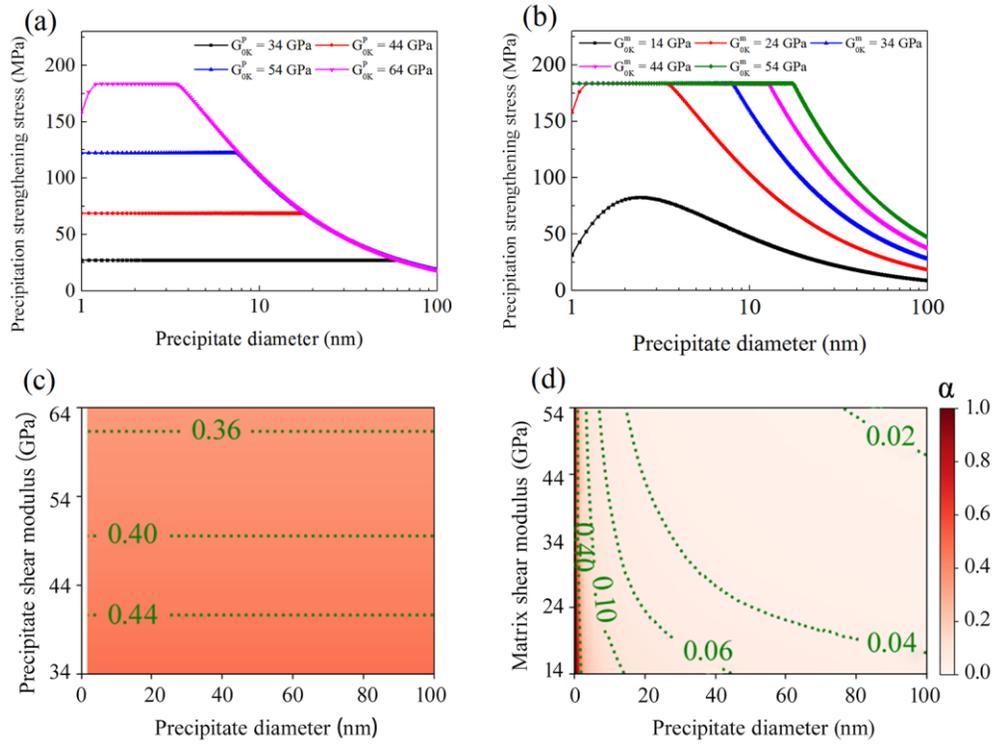

**Figure 7. Effect of shear modulus on precipitation strengthening.** (a), (c) The effects of the precipitate's shear modulus on precipitation strengthening stress and the thermal activation factor. (b), (d), The effects of matrix's shear modulus on precipitation strengthening stress and the thermal activation factor.

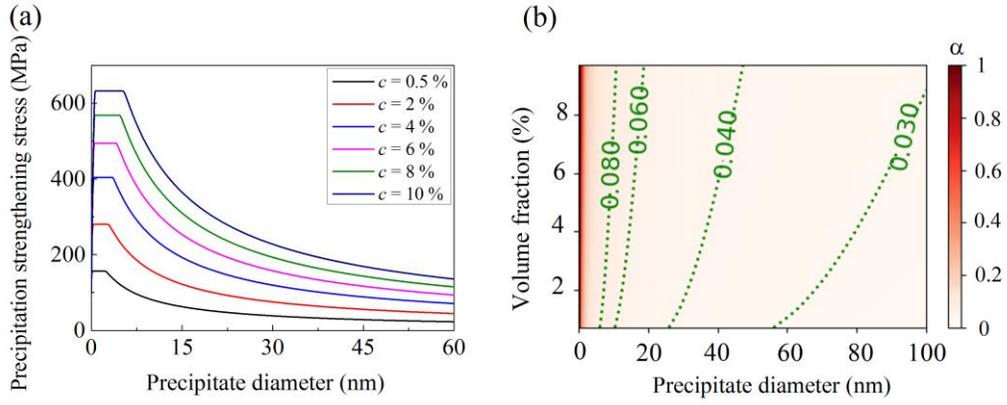

**Figure 8. Effect of volume fraction on precipitation strengthening.** (a) The effects of volume fraction on precipitation strengthening stress. (b) The effects of volume fraction on thermal activation factor.

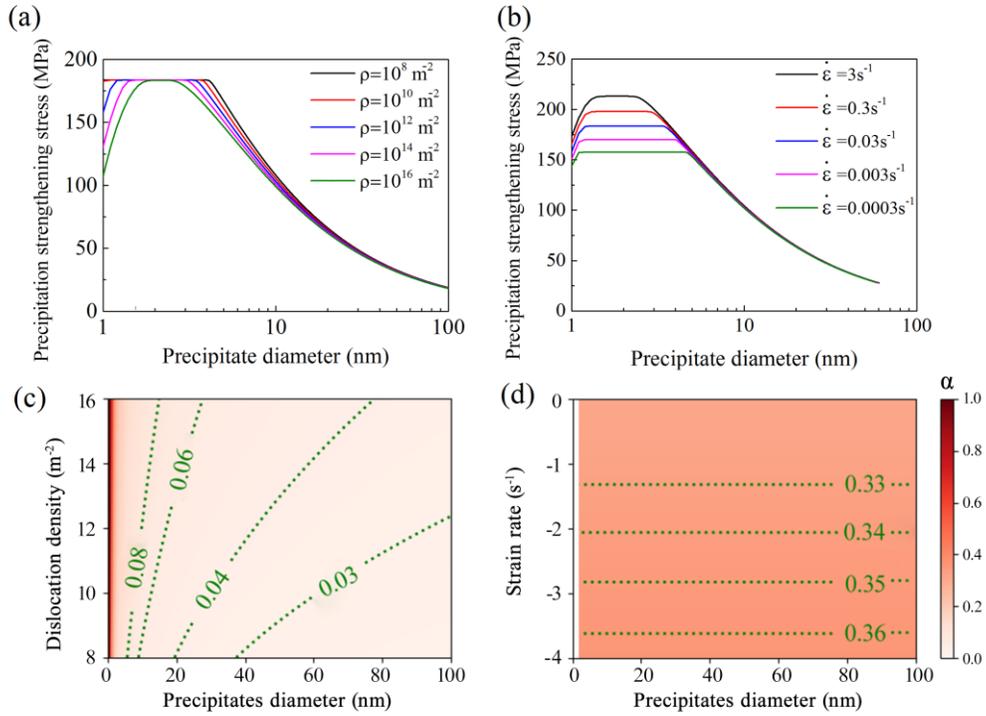

**Figure 9. Effect of dislocation density and strain rate on precipitation strengthening.** (a), (c) The effects of matrix mobile dislocation density on precipitation strengthening stress and thermal activation factor. (b), (d) The effects of strain rate on precipitation strengthening stress and thermal activation factor.

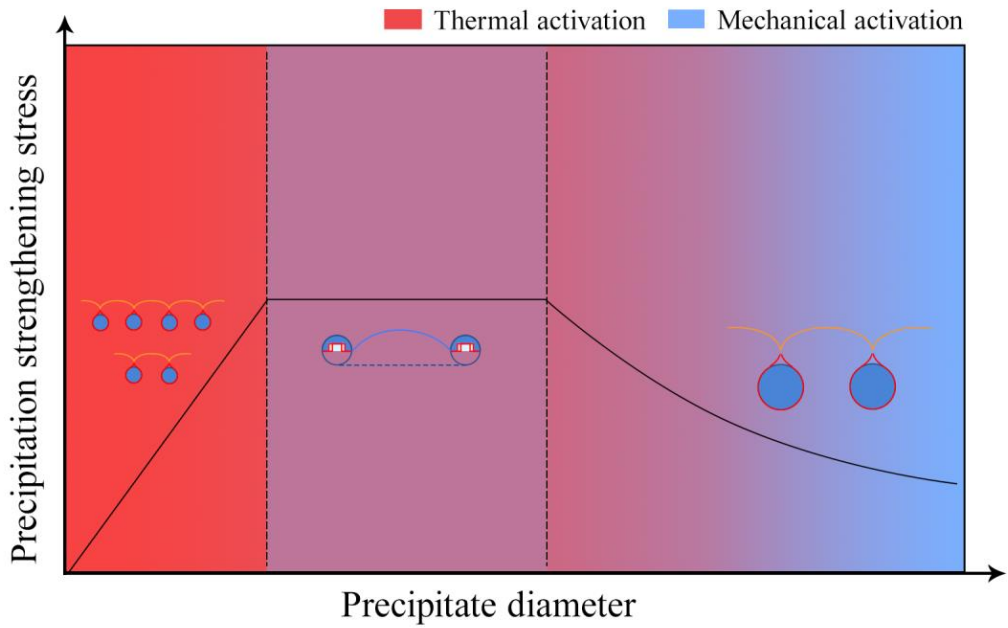

**Figure 10. Schematic of thermally and mechanically activated contributions to precipitation strengthening.**

Table. 1 Calculated parameters for precipitation strengthening model.

| Parameters | AlSc | MgAlScZr | AlScZr | FeCrNiAlV |
|---|---|---|---|---|
| $\dot{\varepsilon}$ (s$^{-1}$) | $3 \times 10^{-2}$ | $10^{-4} \sim 10^{-2}$ | $1 \times 10^{-2}$ | $1 \times 10^{-3}$ |
| $G_m$ (GPa) | 25.4 | 26 | 25 | 46 |
| $b_m$ (nm) | 0.286 | 0.286 | 0.286 | 0.249 |
| $\rho_m$ (m$^{-2}$) | $10^{13}$ | $10^{13}$ | $10^{13}$ | $10^{13}$ |
| $\lambda_m$ ($b$) | 100 | 100 | 100 | 100 |
| $\omega_m$ (HTz) | 3.5 | 3.5 | 3.5 | 3.5 |
| $\nu_m$ | 0.34 | 0.34 | 0.35 | 0.30 |
| $c$ (%) | 0.75 | 0.46 | 0.3~0.5 | 20 |
| $G_p$ (GPa) | 67.9 | 78 | 73 | 61 |
| $b_p$ (nm) | 0.289 | 0.289 | 0.289 | 0.249 |
| $\rho_p$ (m$^{-2}$) | $10^{16}$ | $10^{16}$ | $10^{16}$ | $10^{16}$ |
| $\lambda_p$ ($b$) | 10 | 10 | 10 | 10 |
| $\omega_p$ (HTz) | 3.5 | 3.5 | 3.5 | 12.1 |
| $\nu_p$ | 0.17 | 0.21 | 0.21 | 0.31 |